\documentstyle [preprint,prl,aps]{revtex}

\begin{document}
\title{Broken Symmetry, Boundary Conditions, and Band Gap Oscillations in
Finite Single Wall Nanotubes}
\author{Lei Liu$^1$, C. S. Jayanthi$^1$, H. Guo$^2$, and S. Y. Wu$^1$}
\address {$^1$Department of Physics, University of Louisville,
Louisville, Kentucky 40292}
\address{$^2$Center for the Physics of Materials and 
Department of Physics, McGill University, Montreal, PQ, Canada H3A 2T8}
\date{\today}
\maketitle

\begin{abstract}

We have shown that the interplay between the broken symmetry of finite
single-wall nanotubes (SWNT) and the boundary conditions affects the electronic 
properties of SWNTs in a profound way. For finite SWNTs $(p,q)$ characterized 
by $p=k+l$, $q=k-l$, $p-q=2l$, $l=0,1,\cdots,k$, and $k=1,2,\cdots$,
we found that the band gaps of finite SWNTs belonging
to a certain $k$ exhibit similar well-defined oscillating patterns
but with diminishing amplitudes from the armchair $(l=0)$ to the zigzag
$(l=k)$ SWNTs. These profound changes hold intriguing implications in 
the potential utilization of these finite NTs as the basic component
of molecular scale devices.
\end{abstract}


Very recently, there is a series of interesting studies related to
the feasibility of utilizing carbon nanotubes (NTs) of finite 
lengths as the basic component for molecular scale electronic 
and magnetic devices
\cite{Dekker98,Aouris98,Ago99,Antonov99,Soh99,Dekker99,Dekker00}.
Specifically, there have been experimental evidences of   
increasing band gap in NTs of decreasing length 
\cite{Dekker97,Dekker97a}. Theoretical calculations 
\cite{Avouris99} have shown that an armchair single-wall NT (ASWNT), while it 
is metallic when infinitely long, 
developed a band gap when it is short and this band gap exhibited a 
well-defined oscillation as a function of the length of the NT. 
Mehrez {\it et al}\cite{Guo00} found in a theoretical study different 
transport behavior for ASWNT-based magnetic tunnel junctions of 
different lengths. These studies had demonstrated that the finiteness 
of ASWNT has altered the fundamental properties of its electronic 
structure. However, the physics underlying these profound changes has 
not been delineated. To explore the full potential of NTs as molecular 
scale devices, it is imperative to first determine the nature of the 
electronic structure of all types of SWNTs of finite length, and
then to have a clear understanding of its underlying physics.
In other words, the logical question to be raised is whether these
or any other profound changes are to be found for finite SWNTs of
general chirality. In this Letter, we address this issue in 
terms of the symmetry breaking and the boundary condition.

For an infinite graphene sheet, each atom in the bonding network
forms three $\sigma$-bonds with its three equivalent neighbors. 
In addition, the remaining itinerant electron associated with the atom 
may form a $\pi$-bond with the itinerant electron of any one of the
three neighbors, resulting in the formation of a double bond between 
the atom and that particular neighbor. Because of the equivalence 
of the three neighbors due to the symmetry of the infinite graphene 
sheet, there are in fact three such equivalent double bond 
distributions (see, for example, the pattern displayed in Fig. 1). 
These three bonding configurations are in resonance, thus leading 
to a network of equivalent resonant bonds for an infinite graphene 
sheet with no manifestation of any of the double bond patterns. 
When the infinite graphene sheet is rolled up into an infinite 
nanotube, nothing is changed as far as the bonding pattern is 
concerned. The situation is, however, entirely different for a 
nanotube of finite length. In this case, the double bond pattern 
can survive under appropriate conditions. The scenario can be 
understood as follows. Fig. 1 shows one of the possible double 
bond distribution. If the carbon atoms in the first section (row) 
are replaced by hydrogen atoms, an examination of the pattern in 
Fig. 1 indicates that that particular double bond distribution will 
survive if and only if the sheet is terminated by replacing the carbon 
atoms in the $3m^{th}$ section ($m=1,2,\cdots$) by hydrogen atoms. 
In this situation, there are $3m-2$ sections in the finite graphene 
sheet sandwiched between the top and bottom sections of passivating 
hydrogen atoms. In other words, a finite graphene sheet with 
"lengths" of $3m-2=3n+1$ sections ($n=0,1,2,\cdots$) can sustain 
the double bond distribution, indicating the formation of the 
$\pi$-bond that in turn results in a large gap separating the 
bonding $\pi$ state with the antibonding $\pi^*$ state. On the other 
hand, for finite graphene sheets with lengths of $3n$ and $3n-1$ 
sections, no characteristic double bond distribution of the graphene 
sheet can be maintained and hence only small gaps exist.  
When the finite graphene sheet is rolled up into a tube with its 
circumference along the horizontal direction as shown in Fig. 1, a 
finite ASWNT is obtained. The analysis of the broken symmetry 
associated with the {\it finiteness} of the 
graphene sheet and the boundary 
condition imposed on the finite sheet immediately leads to the 
conclusion that ASWNTs with lengths of $3n+1$ sections will 
possess large band gaps while ASWNTs with lengths of $3n$ or $3n-1$ 
sections will only have relatively small gaps. Thus it is the
existence or non-existence of the double bond configuration as a 
result of the interplay between the broken symmetry associated with 
the {\it finiteness} of the ASWNT and the boundary condition that is 
responsible for the well-defined oscillatory behavior of the 
band gap of ASWNTs reported in \cite{Avouris99}.

To demonstrate our scenario of bonding patterns for finite ASWNTs 
of different lengths, we calculated the bond charge distribution 
for these ASWNTs, using the extended H\"{u}ckel molecular orbital 
(EHMO) method\cite{Landrum} as it has been shown to give 
qualitatively correct result in comparison with ab initio 
calculations\cite{Avouris99}. In the calculation, the C-C and C-H 
bond lengths were kept at commonly accepted values of 
{1.42\AA} and {1.09\AA} respectively. The left panel of Fig. 2 gives 
the bond charge of the three different types ($3n+1$, $3n-1$, and $3n$)
of (6,6) finite ASWNTs vs the bond label. Here, the 
bonds in Fig. 1 are labeled by integers continuously from the 
top to the bottom. For example, bonds connecting atoms in the 
first section (hydrogen) and those in the second 
section (carbon) are labeled by 1, bonds connecting carbon 
atoms within the second section by 2, bonds connecting carbon 
atoms between the second and third section by 3, 
and so on (see Fig. 1, note that bonds labeled by $3s+2$ are 
double bonds, with $s=0,1,\cdots$). 
From the left panel of Fig. 2, it can be seen that, 
for the finite ASWNT with a length of 16 ($3n+1$ type, with $n=5$) 
sections, the bond charge distribution exhibits the 
characteristic pattern of the double bonds of a graphene sheet as shown in Fig.  1. 
The existence of the double bond configuration indicates the 
formation of the $\pi$-bonds, thus leading to a large band 
gap between the bonding $\pi$ state and the antibonding $\pi^*$ 
state. On the other hand, for finite ASWNTs with lengths of 14 
($3n-1$ type with $n=5$) sections and 18 ($3n$ 
type with $n=6$) sections, the bond charge distribution does not 
follow the correct double bond pattern. Hence ASWNTs with lengths 
of 14 ($3n-1$ type) and 18 ($3n$ type) sections 
can only have relatively small gaps. We have also used the EHMO method 
to calculate the band gap of the finite (6,6) ASWNT as a function of 
its length. The result is displayed in Fig. 3 for finite ASWNTs 
with even number of sections (to be discussed later), and in the inset 
of Fig. 3 for ASWNTs with both even and odd number of sections. 
The well-defined oscillatory behavior of the band gap as a function of the 
length of the NTs exhibited in both displays shows the direct correlation to 
the bond scenario presented above. This oscillatory pattern 
of the band gap of the finite ASWNT as a function of the length 
anchored by large band gaps of the order of magnitudes of $eV$s 
for lengths of $3n+1$ sections represents a fundamental change in the 
electronic structure of ASWNTs from a 
pseudo-one dimensional metallic conductor when they are infinitely 
long to a zero-dimension large band gap semiconductor at selected lengths.

Having established the pivotal roles played by the finiteness of 
the ASWNT and its interplay with the boundary condition, it will 
be interesting to find out how these factors or any modification 
of them affect the electronic structure of a finite SWNT of 
general chirality. The totality of SWNTs $(p,q)$ can be divided 
into two groups. Group I is characterized by $p=k+l$, $q=k-l$,
$l=0,1,2,\cdots,k$, and $k=1,2,\cdots$, with $p-q=2l$ while 
group II by  $p=k+l$, $q=k-l-1$, and $l=0,1,2,\cdots,k$, 
with $p-q=2l+1$. In this Letter, we focus our study on finite 
SWNTs belonging to Group I. Specifically, we chose $k=6$ to demonstrate the 
result of our study. The family of NTs in this subset includes the 
armchair NT (6,6) (metallic when infinitely long), 
NTs (7,5), (8,4), (10,2), (11,1) 
(semiconducting when infinitely long), 
the NT (9,3) (semimetal when infinitely long), 
and the zigzag NT (12,0) (metallic when infinitely long). 
We started our study by mapping the (7,5) NT onto a flat surface 
as shown in Fig. 4. Imagining that the carbon 
atoms in the top section of the NT are passivated by a section of 
hydrogen atoms (not shown in Fig. 4), hydrogen-passivated 
finite (7,5) NTs of different lengths can be 
obtained by terminating the sheet at different sections at the 
other end with hydrogen atoms. A close examination of Fig. 4 
reveals three interesting properties for (7,5) NTs of 
finite length. (i) Similar hydrogen-terminations at both ends of 
the finite (7,5) NTs can only be achieved for such NTs of even 
number of sections. Referring to Fig. 4, the first section of the
(7,5) NT is outlined by the carbon atoms with dangling bonds 
(denoted by integer 1). It can be seen that there are two steps in the 
first section where the adjacent atoms are mis-aligned. 
While each carbon atom in the first section is connected 
to one passivating hydrogen atom at the top (not shown 
in Fig. 4), the corner atom, one at each step, is connected to 
two atoms in the second section (denoted by
integer 2). On the other hand, 
every atom in the second section is connected to one atom in the 
third section (denoted by integer 3). This alternating pattern persists throughout the 
sheet from the top to the bottom. As a result, different boundary conditions 
exist at the top and at the bottom for (7,5) NTs with odd number of sections, 
while similar boundary conditions exist at both ends of finite (7,5) 
NTs with even number of sections. In fact, this scenario is 
also true for the other SWNTs in group I except ASWNTs $(k,k)$. 
From Fig. 1, it can be seen that similar boundary 
conditions exist at both ends for ASWNTs regardless whether they 
contain odd or even number of sections. Hence, in this respect, 
the finite ASWNT is the exception rather than 
the rule among all the finite NTs in group I. (ii) Figure 4 
shows that, for (7,5) NTs with 
even number of sections, the double bond pattern (with some 
"defects" to be discussed in (iii)) can be maintained for 
those containing $3n+1$ sections (with odd $n$). Such pattern, 
however, does not exist for those containing $3n-1$ (with odd $n$) 
sections or $3n$ (with even $n$) sections. This picture is 
consistent with the situation discussed for the ASWNT, as 
there are again the same three types of finite NTs with the same 
identical properties. It also means that a well-defined 
oscillatory pattern consistent with the rule characterizing 
the oscillatory behavior of the ASWNTs must exist for the band gap 
of (7,5) NTs containing even number of sections as a function 
of the length. (iii) For the (7,5) NTs, the top boundary (first 
section) is characterized by two steps where adjacent atoms are 
misaligned, creating what are akin to two "stacking" faults 
(see Fig. 4). The defects associated with the stacking faults 
are highlighted as hexagons bordered by solid/dash 
bonds. These bonds are resonant bonds formed as a result of 
the resonance between two mis-aligned
equivalent bonding configurations (see Fig. 4).
We have calculated the bond charges of the defect configuration.
We found the bond charges for the resonant 
bonds to lie between those of the single bonds and double bonds.
These findings confirm our picture regarding the nature of the 
defect configuration. The effect of the presence of these 
defects in the double bond pattern is to 
reduce the magnitude of the band gap (see, Fig. 3).

We have calculated the bond charges of (7,5) NTs of various 
lengths. The results for (7,5) NTs of 
length 16 (type $3n+1$ with $n=5$) sections, 14 (type $3n-1$ with 
$n=5$) sections, and 18 (type $3n$ with $n=6$)
sections are shown in the right panel of Fig. 2. 
The results are similar to the case of ASWNT with the bond charge 
distribution for 16 sections exhibiting the double 
bond pattern shown in Fig. 4 while those for the 
other two cases (14 and 18 sections) do not, confirming the 
picture presented in (ii) in the preceding paragraph. The 
consequences of the three properties enunciated above leads to 
the prediction that the band gap of (7,5) NTs with even number of 
sections exhibits a regular oscillatory pattern similar to the one 
for ASWNTs, namely, large gaps for NTs of 
$3n+1$ sections (with odd $n$) and relatively small gaps 
for NTs of $3n-1$ (with odd $n$) 
sections or $3n$ (with even $n$) sections. However, because of 
the presence of the "stacking" 
fault defects, the amplitudes of oscillations are smaller 
compared to the corresponding 
ones for ASWNT. We have also calculated the band gaps for 
the (7,5) NTs with even number of sections. The result 
displayed in Fig. 3 indeed shows that the band gap of 
finite (7,5) SWNTs exhibits similar oscillatory pattern 
as that of the finite ASWNTs but 
with smaller amplitudes of oscillations, just as 
predicted using the three properties 
discussed above.

The argument presented in the discussion of the three properties 
for (7,5) NTs applies also to (8,4), (9,3), (10,2), (11,1) and 
(12,0) NTs. However, this series of NTs will 
have increasing number of defects from (7,5) to (12,0). 
Therefore, similar oscillatory patterns, but with diminishing 
amplitudes of oscillations, are expected for these NTs. The 
result of our calculation shown in Fig. 3 has indeed also confirmed 
this prediction. The same argument is also valid for
all the other families 
of SWNTs in group I. We 
have studied families of NTs in this group for other $k$ values. 
We found similar results as expected. For example, the band gaps 
of the SWNTs in the family corresponding to $k=5$, including 
the series of NTs (5,5), (6,4), (7,3), (8,2), (9,1) and (10,0), 
show exactly the same behavior as that for the family corresponding 
to $k=6$. In particular, there is hardly any difference between 
the oscillating pattern for the zigzag SWNT (ZSWNT) 
(10,0) in the $k=5$ family and that for the ZSWNT (12,0) in the
$k=6$ family. It is interesting to note that both ZSWNTs have 
vanishing gap when longer than 4 sections 
while the former is semiconducting and the later is 
metallic when both are infinitely long. 
The reason for them to have indistinguishable behavior when 
they are short must be attributable to the fact that they 
both have the largest number of "stacking" fault defects 
in their respective family and these defects overwhelm the 
large gap associated with the double bond pattern.

Finally, we have calculated the resistance at the Fermi energy of 
the family of NTs with $k=6$, using the Landauer's 
formula\cite{Datta95,Liu00}. In the calculation, we modeled the 
leads following the procedure of Mehrez {\it et al}\cite{Guo00}. The Hamiltonians 
of the samples were constructed using the EHMO method. The result is shown 
in Fig. 5. It can be seen that the resistance curves for these 
NTs of finite lengths also exhibit a well-defined oscillatory 
pattern. A side-by-side examination of the resistance curves in 
Fig. 5 and the corresponding band gap curves in Fig. 3 reveals a 
close correlation between the gap and the resistance.

In conclusion, we have shown that finite SWNTs in each family
$(k)$ in group I exhibit similar pattern for band gap oscillations
with diminishing amplitudes from ASWNT to ZSWNT.
We have also established that these characteristic 
patterns for the band gap are 
the consequences of the interplay between the broken symmetry 
associated with the 
finiteness of the NTs and the boundary conditions. 
The fundamental change in the 
electronic structure of all the finite SWNTs in group I 
has very intriguing implications for 
utilizing these finite NTs as molecular scale devices. 
For example, our result indicates 
that it is more appropriate to use ZSWNTs as molecular scale 
conductors rather than 
ASWNTs. On the other hand, ASWNTs of selected 
lengths may be used as large band 
gap molecular scale semiconductors. 
Furthermore, the existence of the double bond pattern raises
the possibility of fabricating nanoscale coils \cite{Louie96}.
We have also studied the electronic structure of 
finite SWNTs in group II. It turned out that different 
physics is operational in that 
situation. The result of that study will be presented elsewhere.

This work is supported by NSF through the grant (DMR-9802274). 
We acknowledge the 
many useful discussions with Prof. Guang-Yu Guo and the use 
of the computing 
resources at the University of Kentucky Center for 
Computational Sciences.      
	
\newpage
\noindent{\bf FIGURES}
\vskip 0.1 in
\noindent {Fig. 1 A typical double bond pattern in a graphene sheet.}
\vskip 0.1 in
\noindent{Fig. 2 The (average) bond charge vs the bond label.}
\vskip 0.1 in
\noindent {Fig. 3 The band gap vs the length (the number of sections) for SWNTs 
containing even number of sections corresponding to the subset $k=6$ . The inset shows 
the band gap curve for the (6,6) ASWNTs containing both odd and even number of 
sections. }
\vskip 0.1 in
\noindent{Fig. 4
The mapping of (7,5) SWNTs onto a 
flat surface and the double bond pattern. 
The stacking fault defects are highlighted as hexagons bordered 
by resonant (solid/dash) bonds.}
\vskip 0.1 in
\noindent {Fig. 5 The resistance vs the length for the series of SWNTs (6,6), 
(7,5), and (8,4) containing even number of sections. The inset gives the resistance vs the 
length curve for the ASWNTs (6,6) containing both odd and even number of sections. }

\end{document}